\documentstyle[12pt,aaspp4]{article}

\lefthead{Shapley et al.}
\righthead{Multivariate analysis of spiral galaxy luminosities}

\def\ergcm2s{erg cm$^{-2}$ s$^{-1}$} 
\def\etal{et al.}		

\begin{document}

\title{A Multivariate Statistical Analysis of Spiral Galaxy Luminosities.\\
I. Data and Results}

\author{Alice Shapley}
\affil{California Institute of Technology,\\
Pasadena CA, 91125, USA}

\author{G. Fabbiano }
\affil{Harvard-Smithsonian Center for Astrophysics,\\
60 Garden Street, Cambridge, MA 02138}

\author{P. B. Eskridge}
\affil{Ohio State University,\\
Dept. of Astronomy,\\
140 W 18th Ave, Columbus, OH 43210}

%


\begin{abstract}
We have performed a multiparametric analysis of luminosity data for a sample of 234 
normal spiral and irregular galaxies observed in X-rays
with the {\it Einstein Observatory}.  This sample 
is representative of S and Irr galaxies, with a good coverage of 
morphological types and absolute magnitudes. In addition to X-ray and optical data, we 
have compiled  H-band magnitudes, IRAS near- and  far-infrared, and 6cm radio continuum 
observations for the sample from the literature. We have also performed a careful 
compilation of distance estimates. We have explored the effect of 
morphology by dividing the sample into early  (S0/a-Sab), intermediate (Sb-Sbc), and 
late-type (Sc-Irr) subsamples.  The data were analysed with bivariate and multivariate
survival analysis techniques that make full use of all the information
available in both detections and limits.
We find that most pairs of luminosities are correlated when
considered individually, and this is not due to a distance bias.
Different luminosity-luminosity correlations follow different power-law relations. 
Contrary to previous reports, the $L_X - L_B$ correlation follows a power-law 
with exponent larger than 1. Both the significances
of some correlations and their power-law relations are morphology 
dependent. Our analysis confirms the `representative' nature of our 
sample, by returning well known results derived from previous analyses
of independent samples of galaxies (e.g., the $L_B - L_H$, $L_{12} - L_{FIR}$, 
$L_{FIR} - L_{6cm}$ correlations). Our multivariate analysis suggests that there are 
two fundamentally strong correlations, regardless of galaxy morphology, when all
the wavebands are analyzed together with conditional probability methods.
These are the $L_B - L_H$ and the $L_{12} - L_{FIR}$ correlations.
As it is well known, the former links stellar emission processes, and points to a basic
connection between the IMF of low-mass and intermediate-to-high-mass
stars. The latter may be related to the heating of small and larger
size dust grains by the same UV photon field.
Other highly significant `fundamental' correlations exist, but are
morphology-dependent. In particular, in the late sample (Sc-Irr) we see
an overall connection of  mid-,  far-IR, and radio-continuum
emission, which could be related to the presence of star-forming activity
in these galaxies, while in early-type spirals (S0/a-Sab), we find no strong
direct link of FIR and radio continuum.
This paper gives a compilation of both input data and results of our
systematic statistical analysis, as well as a discussion of potential
biases. Results relevant to both X-ray and multiwavelength
emission properties are analyzed futher and discussed in Paper II.
\end{abstract}




%

\section{Introduction}

Understanding the structure, formation and evolution of galaxies is one
of the main themes of present-day astrophysics. This quest is made
difficult by the complexity of galaxies, their interactions with their
environment, and our limited knowledge of their observational characteristics
(see \cite{gf90}). While most of the studies of galaxies
make use of individual energy bands, chiefly the optical, but also the radio,
and more recently the X-ray and infrared (IR), it is rarer to find work
comparing data from two or more emission windows. Yet, when this is done
interesting insights may follow. For example, the comparison of H-band and B-band 
photometry led to the discovery of the well known color-magnitude relation
for spiral galaxies (\cite{ahm79}; \cite{tma82}), a non-linear correlation between $L_B$
and $L_H$. The comparison of 
IRAS far-IR and radio continuum data led to the discovery of the well-known
strong correlation and to the convincing association of the radio continuum 
emission with the star-forming stellar population (\cite{dic84}; \cite{hel85};
\cite{dej85}); comparison of
CO, H$\alpha$ and IR data led to constraints on star formation efficiencies 
in spirals (e. g. \cite{y90}); comparison of multiwavelength data, including
X-rays, in late-type spirals suggested the prevalence of intrinsically obscured 
compact star-forming regions in higher luminosity galaxies (\cite{fgt88};
\cite{tfb89}).

In this paper we report the statistical analysis of the sample of 234
`normal' spiral and irregular galaxies observed in X-rays with the {\it Einstein Observatory}
(\cite{g79}), as reported in `An X-ray Catalog and Atlas of Galaxies' by \cite{fkt92}
(FKT hereafter). The present works complements the papers on the statistical 
analysis of the 148 E and S0 galaxies from FKT (Eskridge, Fabbiano \& Kim 1995a, b, c)
and completes the statistical analysis of the FKT sample. Previous exploratory
work on spiral and irregular galaxies (\cite{ft85}; \cite{fgt88}, see \cite{f90}), 
was based on a much smaller sample of 51 galaxies.
For the purpose of the present work, we have augmented the data presented in FKT (X-ray 
and optical), with H-band, mid and far-IR (IRAS), and 6cm radio continuum magnitudes 
and flux densities from the literature. This gives us representative
coverage over the entire electromagnetic emission spectrum of spiral galaxies, and
allows us to explore the full range of emission processes and the interaction of
different galaxian emission components.  These phenomena include direct or 
reprocessed stellar emission (optical and IR); emission from the evolved component of 
the stellar population, hot ISM, and nuclei (X-rays); synchrotron emission of cosmic-ray 
electrons interacting with the galaxian magnetic fields,  and thermal emission of 
$\sim 10^4$K hot ISM (radio continuum).  These different emission bands have  different
sensitivities to absorption, and their comparison may also give us some insight on the
dust content of the emitting regions (e.g.\cite{pal85}; \cite{ft87}).

The size of the FKT sample of spiral and irregular galaxies allows us 
to explore the dependence of these processes on galaxian morphology, one of the key
parameter-axis in spiral galaxies (\cite{whi84}). Such a dependence was suggested by earlier work
(\cite{ft85}; \cite{fgt88}), but those results were based on much smaller samples.
Here we analyse separately bulge-dominant (S0/a-Sab), intermediate (Sb-Sbc), and late-type
(Sc-Irr) galaxies; we then intercompare these results and we compare them with those of the
entire sample.  

This is the first paper of a 2-paper series.  In this first paper we describe the sample and the 
data analysis; we report the results of the analysis; and we discuss the possible
effects of selection biases. In the companion paper (Fabbiano \& Shapley 2001; 
hereafter Paper II) we look in detail at the astrophysical significance of the results,
and we compare our results with those of other related work. 

\section{ The Sample }

The sample used for the statistical analysis consists of 234 spiral and 
irregular galaxies belonging to the FKT sample (Fabbiano \etal 1992). As
described in FKT, it consists of relatively nearby galaxies, all
observed with {\it Einstein}. This was the first sample of galaxies ever
to be observed in X-rays, and was mostly assembled to be
a representative (optically selected) sample of normal galaxies, spanning the full 
range of morphologies and luminosities. To reduce selection biases, FKT used the RSA (\cite{sand87})
and RC2 (\cite{RC2}) as basic selection catalogs, by adding to the sample all RSA/RC2 galaxies 
present in the regions of the sky observed with {\it Einstein} included in the catalog.
Fig.~1 shows the histogram of absolute magnitudes of our sample. It compares
well with the corresponding histogram from the RSA.

The FKT sample includes galaxies of all morphological types. 
Fig.~2 shows the distribution of morphologies in the spiral sample. All types from
S0/a ($T = 0$) to Irr ($T = 10$) are represented. For the purpose of our 
analysis, besides considering the entire sample of 234 spiral and 
irregular galaxies, we also divided the sample into three morphological
subsamples:the `early' sample, $T = 0 - 2$ (58 S0/a-Sab, and 7 Amorphous); 
the `intermediate' sample,  $T = 3 - 4$ (Sb-Sbc, 62 galaxies);
and the `late' sample $T = 5 - 10$ (Sc-Irr, 107 galaxies). Since the early sample
in this definition would include 7 Amorphous galaxies (see \S3.), we further 
excluded these galaxies. So defined, these subsamples are representative of
bulge-dominant systems, bulge/disk systems, and disk/arm-dominant systems respectively.
Dividing the sample according to morphology 
is motivated by earlier results which have suggested that the 
multiwavelength statistical properties of spiral galaxies are morphology-dependent 
(\cite{ft85}; \cite{fgt88}). 

The FKT spiral sample includes a number of AGN. Twenty of these are X-ray-bright 
powerful Seyfert galaxies and were identified as such in FKT. 
The nuclear X-ray source in these galaxies totally dominates the X-ray emission,
which is then the expression of the AGN and cannot give us any useful indication 
on the general `normal' X-ray emitting population.
We have excluded galaxies flagged as AGN by FKT   from our analysis, but they are
included in some of the figures. However, more recent work with more sensitive 
data has revealed that nuclear activity, once thought to be an extraordinary 
phenomenon, is instead rather ubiquitous, albeit at a very low level (\cite{hfs97}).
The separation of AGN from `normal' galaxies becomes then a philosophical issue
in the case of low luminosity activity. Since most bulge galaxies may host nuclear
massive black holes ( e.g. \cite{mag98}), undetected nuclear activity is alway possible.
We have retained in our working sample 51 galaxies found by \cite{hfs97}
to have some indication of nuclear activity in their optical spectra. These 
include 19 low-luminosity Seyfert nuclei, as well as LINERs and nuclei with spectra
intermediate between HII regions and LINERs (transition objects).
Typically their nuclear X-ray sources, based on the cases where high enough resolution
is available (e.g., FKT), is just one of several identifiable  components in the (0.2-4)~keV
{\it Einstein} band. More recent ROSAT observations (with 5" resolution) of face-on
spiral galaxies show that near-nuclear relatively bright ($L_X \sim 10^{37-40} \rm ergs/s$)
sources are rather common, but their nature is not clear: they may be low-luminosity AGN
or bright black hole binaries, or bright young SNR (\cite{cm99}).
Therefore, we do not find it justifiable to single out these galaxies.
However, there may be energy bands where these faint nuclei may dominate, and this
is discussed in Paper II.

The FKT sample is neither X-ray selected, nor statistically complete: it is not
volume or flux limited. Therefore it cannot
be used to derive X-ray luminosity functions of spiral galaxies. 
However, as long as the sample is representative of the
range of morphological types, and covers a fair range of galaxy luminosities, it
can be used for studying the relations among different emission bands in galaxies.
To check for possible peculiarities, it is important to compare our results with
analogous results from independent studies, using different `representative' samples 
chosen for different purposes with different criteria. Discrepancies may indicate that one
of these samples may not be indeed representative of the population that it
purposes to study (that of spiral and irregular galaxies), and may indeed suffer from peculiar
selection biases. For this type of comparison it is particularly important to look at the overall
multi-wavelength spectrum of correlations, and see if we retrieve some of the well known 
(non X-ray) results that have been found from separate, independent studies.
This type of comparison is pursued here and is discussed in greater detail in Paper II.
We show there that our results are in agreement with well known IR-optical-radio
relationships in spirals, and that therefore ours is a fair sample for this type of
study.

\section{ The Data }

Table 1 lists the galaxies (including the AGN, which are flagged) ; their coordinates; 
morphological types ($T$); distances; X-ray fluxes; optical($B$) and near-IR ($H$) 
magnitudes; {\it IRAS} and  radio continuum flux densities; and gives the sources for the
entries. In the case of non-detections, $3 \sigma$ upper limits are given.
Notes and  references to Table 1 are given in Table 2: items 1-5 refer to FKT and 
other X-ray references; items 6-17 are references and notes on the infrared data;
items 18-44 refer to the radio continuum data; items 45-68 refer to the H-band data. 
Additional information on the H-band data is given in Table 3.
The variables used for the 
statistical analysis consist of the logs of the luminosities calculated from Table 1, 
and are listed in Table 4.

Details on Table 1 and on the derivation of Table 4 entries follow:

\noindent
\underline {Type ($T$)}.
The galaxies in our sample range in morphological type 
from $T = 0$ to $T = 10$, corresponding to Hubble types from S0/a to Irr, as listed
in FKT.  The sample also includes 7 T=0 galaxies with irregular morphology. 
These are indicated by an `A' (Amorphous; \cite{sand87}) in the T column.

\noindent
\underline {Distance (D)}.
We have performed a thorough literature search for distance information for our
sample.  Thus the distances in Table 1 differ from those in FKT, which were derived 
from Tully (1988) for $H_0 = 50 {\rm\:  km \: s}^{-1} {\rm \: Mpc}^{-1}$.
Details are given in Appendix A.1.

\noindent
\underline{X-ray flux ($f_{X}$)}.
X-ray data (0.2 - 4.0 keV fluxes or 3$\sigma$ upper limits)
were taken from FKT. 

\noindent
\underline{ Optical magnitudes (B)}.
Optical, extinction and inclination corrected,  ($B$-band) magnitudes are from the 
{\it Third Reference Catalogue of Bright Galaxies}
(RC3; \cite{RC3} ).  They were converted to fluxes
in the B band, following \cite{all73}:  $f_B = 10 ^{ -0.4 \times B - 8.17} \times  990$.

\noindent
\underline{ Near-infrared  1.65 $\mu$m magnitudes (H)}. 
To obtain near-infrared ($H$-band, 1.65 $\mu$m) data for as many galaxies in the
sample as possible, we looked in the {\it Catalogue of
Visual and Infrared Photometry of Galaxies from $0.5 \mu m$ to $10 \mu m$}
(\cite{dev88}), which contains near-infrared measurements
of galaxies  and references from the literature from 1961 - 1985. We found
photometry data for 159 {\it Einstein} galaxies, (140 normal, and
19 flagged as AGN) from the references listed in the {\it Catalogue}.

The $H$-band data were collected from a number of different
sources in the literature, and therefore the idiosyncracies of the
various sources of data needed to be reconciled.
First, different aperture-to-diameter ratios were used
for various galaxy measurements--i.e. a smaller aperture-to-diameter
ratio samples a smaller fraction of the galaxies total
near-infrared magnitude. Also,
several near-infrared filter systems
are represented by the full set of $H$ measurements.
These systems have slightly different zero-points
for the conversion from magnitudes to fluxes and slightly
different central wavelength and bandwidths. 
Since the differences in aperture-to-diameter
ratio and filter system cause systematic
offsets among the near-infrared data and tend to increase
the scatter in correlations, the data must be corrected
before it can be used for statistical analysis.

To correct the data to a consistent aperture system,
we turned to the work of Tormen and Burstein 1995. 
In an effort to recalibrate the near-infrared Tully-Fisher relationship,
Tormen and Burstein normalize a dataset of $H$-band aperture magnitudes
from 1731 galaxies collected over a ten year period by Aaronson and collaborators.
The central problem of homogenizing the datasets consists of correcting
the $H$-band magnitudes to the same aperture/diameter ratio, such that
$log(A/D)=-0.5$. In order to perform this correction, Tormen and 
Burstein determine empirical curves of growth for four different
morphological subgroups, and use the morphologically appropriate
curve of growth to correct the aperture photometry of each galaxy
to the fiducial value of $H_{-0.5}$ (which is the value of
$H$ evaluated at $log(A/D)=-0.5$). We found corrected
$H$-band magnitudes for 87 {\it Einstein} galaxies in Tormen
and Burstein, and adopted these magnitudes as the normalized
near-infrared magnitudes. 

Additionally, there were 72 {\it Einstein}
galaxies for which we found $H$-band data in the 
{\it Catalogue of Near-Infrared and Visual Photometry}, but
which are not included in the Tormen and Burstein sample.
To correct the $H$-band magnitude for these 72 galaxies
in a manner consistent with that of Tormen and Burstein,
we found the isophotal diameter of each galaxy in the RC3 
(corrected for galactic extinction in the same way that
Tormen and Burstein correct the diameter);
we then computed its $log(A/D)$ value based on the RC3 isophotal diameter
and the aperture listed in the literature for the $H$-band measurement; finally,
 we applied
one of the four Tormen and Burstein growth curves, based on
our determination of the galaxy's morphological type, to correct the
listed aperture measurement to the fiducial aperture magnitude for
$log(A/D)=-0.5$--i.e. $H_{-0.5}$. 

In order to check the validity of our method for correcting the magnitudes of these
72 galaxies, we also applied the method to the 87
galaxies included in the Tormen and Burstein paper,
for which we also have uncorrected aperture photometry from the literature.
We wanted to ascertain that our application of the Tormen and
Burstein growth curves gave us corrected values consistent
with the values Tormen and Burstein determined. Indeed, we found
very good agreement between the corrected $H_{-0.5}$ magnitudes
we calculated and the values listed in Tormen and Burstein (fig.~3).

We also addressed the issue of Galactic extinction. Tormen and
Burstein correct all growth curve-corrected magnitudes for galactic
extinction, using the correction $A_H=0.1*A_g$, which usually results
in a correction of less than 0.05 magnitudes. Therefore, the 87 galaxies
in our sample which were also in the sample considered by Tormen and
Burstein have H-band magnitudes which are corrected 
for galactic extinction. We then considered
the 72 galaxies in our H-band sample which were not included in
the Tormen and Burstein paper.  Since H-band magnitudes for these galaxies
were assembled from a variety of sources in the literature, it
was necessary to check whether or not each literature 
source included a correction for galactic extinction. 
We found that for all but two galaxies, the
H-band magnitude in the literature was either corrected for galactic
extinction, or uncorrected but with a required correction of less than 0.05
magnitudes. Therefore, we only added our own corrections to the two
galaxies which did not meet the above stated criteria, IC~342, which
required an H-band correction of 0.30 magnitudes, and NGC~6951, whose
required correction was 0.09 magnitudes.
We did not apply the negative internal extinction correction,
discussed by \cite{wil96}, as
it is clear that this correction is neither significant nor perhaps
even valid in most cases (see the last paragraph, p. 488 of the Willick
paper, where they say that they can't rule out $C^H_{int}=0$ for
Aaronson's H-band data.) The correction is:
$H_{corrected}= H - C^H_{int}$*log(axial ratio)
so that if $C^H_{int}=0$, the internal extinction correction is 0.

Once we corrected all of the $H$ magnitudes to $log(A/D)=-0.5$, 
and had taken into account Galactic extinction, we then converted
each corrected magnitude to an H-flux $(\nu F_{\nu})$ 
(units are ergs sec$^{-1}$ cm$^{-2}$), according to the specific
photometric system used in the reference from which we obtain the measurement.
This conversion requires the $\lambda _{eff}$, the effective central wavelength
of the $H$ filter used, as well as $F_{\nu}(0)$, the $F_{\nu}$ corresponding
to $H$ = 0.0 mag. Therefore, the conversion to H-flux consists of the following:

\begin{equation}
F = \frac{c}{\lambda_{eff}} \times F_{\nu}(0) \times 10^{\frac{-H}{2.5}}
\end{equation}

\noindent where $F$ is the H-flux and $c = 3 \times 10^{10}$ cm sec$^{-1}$
is the speed of light.

Table 3 lists the many conversion systems we used, and the references to which
 they apply.
The reference numbers refer to the system of Table 2.

\noindent
\underline{{\it IRAS} flux densities ($f_{\nu(12)}, f_{\nu(25)}, f_{\nu(60)}, f_{\nu(100)}$)}.
The {\it IRAS} flux densities or 3$\sigma$ upper limits were assembled from several sources
(see Refs. and Table 2). For nearby extended galaxies, we adopted the values reported in
Rice et al. (1988). We obtained 12, 25, and 60 $\mu$m
fluxes for 238 galaxies(218 normal, 20 AGN), 
and 100 $\mu$m fluxes for 237 galaxies (217 normal, 20 AGN).
To derive fluxes from the flux densities,
the {\it IRAS} data were multiplied by the appropriate bandwidths
and normalizations, indicated in the {\it IRAS Explanatory 
Supplement} (\cite{bei84}). To calculate the far-infrared flux $F_{FIR}$,
we followed \cite{lon87}.

\noindent
\underline{6 cm radio continuum ($f_{6cm}$)}. Our literature search yielded 153
flux densities and upper limits (136 for normal galaxies, and 17 for AGN).
We multiplied the radio measurements by a 1$\%$ bandwidth (50 MHz),
to convert flux densities to fluxes.
Previous work on spiral galaxies established the connection between
the non-thermal radio continuum emission of spiral galaxies and
star formation (\cite{fgt88}; \cite{dic84}; \cite{hel85};
\cite{dej85}), making use of 20cm flux densities, 
which are likely to be less contaminated by thermal 
emission (see \cite{ggk82}), and therefore are more representative of the nonthermal 
continuum. The present use of 6cm flux densities was motivated by our desire 
to compare the properties of early-type bulge-dominated spirals
with those of E and S0s (\cite{efk95}). Although the 6cm 
flux cannot be used to prove cleanly the connection betweem cosmic ray
production and star formation, this connection has already been proved (see refs. above). 
Any general consideration about connections of the overal radio emission and 
other galaxian properties will still be valid.

\section{Distributions of $L_X$ and $L_X / L_B$}

Figs. 4  and 5 show the distributions of X-ray luminosities $L_X$ and X-ray-to-optical
ratios $L_X / L_B$ (bright AGN excluded) for the total sample, the three subsamples, 
and E and S0 galaxies (from FKT, \cite{efk95}) for comparison. We see the already 
noticed effect (e.g. \cite{f90}) that the distributions of $L_X$ and $L_X / L_B$
of E and S0 galaxies extend to higher values than do those of spirals. We do not
see any major differences in comparing the three spiral subsamples, with the exception that
the luminosity distribution of $T = 3 - 4$ galaxies does not include any detections
in the lower luminosity bins which are populated in the other subsamples.
However, the distribution of $T = 3 - 4$ limits is consistent with
the presence of less X-ray luminous galaxies.

\section{Correlations}

Fig. 6 displays the scatter diagrams from the fifteen
pairs of luminosity variables under consideration. Several features
of these plots are apparent without any formal statistical analysis.
First, in the plots which feature $L_X$ as the dependent variable,
the flagged AGN lie clearly above the distribution of normal 
spiral galaxies in the vertical direction, indicating the
excess nuclear X-ray emission from these objects. 
Second, most of the pair-wise relationships display
more scatter in the early-type ($T = 0 - 2$, S0/a-Sab) subsample. 
The 7 Amorphous galaxies in the early subsample are indicated by
different symbols in the scatter diagrams. They were not included in the
analysis of this sample. 
Third, for the majority of the luminosity-luminosity
pairs, the distribution of points in the middle ($T = 3 - 4$)
morphological range is basically coincident with the upper
right-hand portion of the distribution of late-type ($T = 5 - 10$)
points. 

Fig. 7 displays scatter plots for luminosity-ratio pairs.
Also here we find that trends are visible in total and late/intermediate
samples, but tend to disappear in the early sample.

We performed bivariate correlation tests and regression analysis
as well as multivariate analysis on these data. 
All information (both detections and limits) was used in the analysis,
by applying survival analysis techniques.
Bivariate analysis was conducted with ASURV Rev 1.1 
(\cite{lav92} and refs. therein), a software package that implements methods of univariate
and bivariate survival analysis (both correlation tests and regression methods).
We tested for the significance of each correlation, and we derived regression
parameters for each of them.
Multivariate analysis addresses the question: is a given correlation intrinsically
significant (and thus indicative of an astrophysical effect), or is it 
the secondary effect of other more fundamental links?
To test for the presence of intrinsic correlation among two variables, 
that would be present even if all other variables did not vary,
we used the Spearman partial rank method
(\cite{ken76}; see \cite{fgt88}, \cite{efk95} for previous applications).
The partial rank analysis takes full advantage of the 
multi-wavelength nature of our set of data and correlations,
providing information that a simple bivariate correlation
analysis cannot supply.
We used the generalized Spearman's rho method from ASURV
to generate correlation coefficients to use in the Partial Rank analysis.

These methods and the results
of the analysis are described in \S6. and \S7.
Below we discuss biases that may affect correlation 
studies and show that our results are free from serious effects.

Distance biases (chiefly the Malmquist bias) are a well known danger
in any correlation analysis, and may result in spurious luminosity
correlations when working with flux limited samples. 
Our results directly confirm that a Malmquist bias is not significant.
First, most regression bisector slopes (see fig.~6 and \S6.) indicate  non-linear
relationships between variables. If the correlations were due to
a Malmquist bias, they  would only appear as  linear relationships
in the $\log-\log$ plane (power-law $\alpha = 1$). 
Second, even for linear correlations, a correlation is evident in 
flux-flux plots (not shown).

Moreover, the characteristics of our sample selection, and the inclusion
of limits in the analysis,  protect us from these effects. 
The {\it Einstein} sample of spiral galaxies contains an optical selection criterion,
but is not defined by any {\it a priori} X-ray flux
or volume limit, and by including upper limits in our
analysis in the X-rays and in the other wave-bands, we have avoided 
the problem of an {\it a posteriori} flux-bias towards higher-luminosity 
objects in the various luminosity parameters.
Censored analysis tools make full use of {\it both detections
and limits}. Under these circumstances, working with fluxes may
provide erroneous results, which are absent when luminosities are used.
(as rigorously demonstrated by \cite{fb83}, see \cite{ft85}, \cite{fgt88}). 
Furthermore, we have used the  Partial Spearman Rank test, to
directly test if a given correlation could have arisen solely from
a distance effect, by including the distance among the variables tested
(\S7. and Appendix A.3). All the bivariate correlations are still very 
significant when the correlation is tested under the hypothesis that
the distance be held fixed, and the results of the multivariate analysis 
are only minimally affected.

Fig.~8 supports our conclusion that the sample is not affected by a 
distance-limited issue: we have a fair sampling of both detections 
and limits at any given distance. 

Correlations cannot be created by a distance bias in our sample; however, 
the presence of upper limits could in some cases imply that we are not in
the presence of a very tight functional relation, but of a `wedge' effect.
Although this possibility cannot be completely discounted, it would not 
change the results of the presence of correlations, it may only weaken
any model based on intrinsic underlying power-laws.

Another distance-related problem consists of the uncertainty in the adopted distance
for any individual galaxy.
Our results are robust to uncertainties in the assumed distances.
We obtain very consistent results when we use directly the set of distances
in FKT, or the present set of Table~1. The FKT distances are mostly 
from \cite{tul88}, corrected for an $H_o = 50$. Some of these distances give
values for nearby galaxies (e.g. M82), which differ significantly from 
recent Cepheid-based estimates. However, these differences do not affect
the results of the correlation analysis. Moreover, we tested the robustness
of our results by randomly perturbing each adopted galaxy distance by either
a factor of two high or low.
This is the outer envelop in the dispersion from a comparison of
distances from galaxian indicators and distances from the Hubble flow that 
we have assembled here (Appendix 1). Even in this extreme case, the basic
correlation slopes stand. Uncertainties arising from different Hubble flow
corrections are much smaller (see Appendix~1, where we compare
YTS and CMB corrections). Comparing runs of our bivariate probability 
and regression analyses for the entire set of correlations using the 
two set of distances shows that in all cases
the resulting effects on the correlations are insignificant (well within the
errors).  The reason is that the cosmic scatter of galaxian properties at a
given luminosity is much greater than the scatter introduced by current
distance uncertainties.

Another possible bias consists of beam-size effects, which could turn
a linear power-law relation between two variables into a non-linear relation,
if one of the variables is observed with a small beam. This effect occurs if the galaxies
further away are systematically more luminous,  of course of smaller angular size,
and therefore not so undersampled by a small beam-size as a nearby galaxy would be. 
A beam-size effect could also obscure the strength of an observed correlation, by introducing
extra scatter into a distribution of points, because the small beam samples a different
fraction of the total galaxy luminosity based on the angular size of the galaxy.

Beam-size effects should not be a problem with the X-ray flux data, since the 
{\it Einstein} field is much larger than any of the galaxies observed, and a method
akin to surface photometry was followed to derive the fluxes, while limits were
derived from areas comparable to the optical extent of the galaxies (see FKT).
Beam-size effects are also not a problem for optical ($B$) and near-IR ($H$) data, 
since in both cases the magnitudes  refer
to the same fraction of the total galaxy. We addressed the finite
nature of the {\it IRAS} beam-size by using \cite{ric88} fluxes, computed
specially for large optical galaxies. To investigate possible beam size dependencies
in the  $6 cm$ data  we have plotted galaxies of different optical sizes with different
symbols in a $L_X - L_{6cm}$ scatter plot (fig. 9). We do not find any significant
differences that may be linked to the galaxy size and conclude that the 6cm data do
not suffer significantly of beam size bias.

Because of the finite resolution of the observations, expecially in the infrared and
X-ray bands, in a very few cases of close-by or interacting galaxies the fluxes may include
the contribution of more than one object. Table 2 shows that, of the galaxies used
for the analysis, no `early' sample galaxy is thus affected, and only 1 (out of 62)
`intermediate' sample galaxy, and 7 (out of 107) `late' sample galaxies suffer of 
source confusion in the IR; given the uneven data coverage, only 3 of these latter galaxies
were included in the multivariate analysis. Inspection of FKT shows 
that confusion in the X-rays is also likely. Given the small percentage of the sample 
suffering of this problem, we do not think that our results would be significantly 
affected. This effect may result in some scatter in the correlations, 
which are however expecially tight in the `late' sample. The only foreseeable effect would 
be to worsen somewhat correlations involving the IR or the X-ray band and one of the other
variables. However, the resulting scatter would be well within the observed dispersion
of the correlations.

Finally, we checked that uncertainties in the $H_{0.5}$ magnitude corrections (\S3.)
did not affect our results, by rerunning the analysis for a set of $H_{0.5}$ values we
calculated using Tormen \& Burstein (1995) prescription (see fig.~3), and comparing 
the results with those obtained from the values in Table~1. The results are virtually identical.

\section{Bivariate Analysis}

We report below the results of  bivariate correlation tests and regression analysis 
for each of 15 luminosity pairs in the matrix of combinations
among the six variables $L_X$, $L_B$, $L_H$, $L_{12}$, $L_{FIR}$, and $L_{6cm}$.
After applying the same tests to correlations including each of the IR variables
($L_{12}$, $L_{25}$, $L_{60}$, and $L_{100}$) we concluded that 12 and 25 $\mu$m 
behave similarly and so we adopted $L_{12}$ as representative of the mid-IR
emission; the same is true for $L_{60}$, and $L_{100}$ and $L_{FIR}$ (which is
a combination of the two).
In addition, we report the results of correlation tests applied to
the X-ray-optical ratio, $L_X/L_B$ and
five other luminosity ratios: $L_{60}/L_B$, $L_{6cm}/L_B$,
$L_{12}/L_B$, $L_{60}/L_{100}$, and $L_H/L_B$.

The bivariate package, BIVAR, in  ASURV 
provides three methods for testing for the
presence of a correlation between two variables containing censored 
data points: the Cox hazard model, the generalized Kendall's tau, and 
the Spearman's rho. Cox's hazard, a parametric method --i.e.
one that requires certain assumptions with respect to the
underlying distribution of the sampled data points-- can
only be used when there is one type of censoring (upper
or lower limits), and when the censoring only occurs
in the dependent variable. The other two methods,
Kendall's tau and Spearman's rho, are non-parametric
tests, operating on the basis of the sample values alone,
without any assumptions regarding the underlying population.
Both of the non-parametric tests can handle censoring in both
the independent and dependent variable. Since many of the
luminosity pairs under consideration contained upper limits
in both variables, we could not apply the Cox method to these cases,
and simply used the Kendall and Spearman correlation tests.
Wherever applicable, the Cox methods gives results -- not shown --
consistent with those of the other two methods.

Table 5 displays the results of the bivariate
luminosity correlation tests for the total ($T = 0 - 10$), early ($T = 0 - 2$), 
intermediate ($T = 3 - 4$), and late ($T = 5 - 10$) samples.
For each test pair and sample, are listed: the number of data points ($N_{tot}$);
the number of upper limits ($N_{lim}$), in the order: limits on the first variable
of the pair, limits on the second variable, and limits on both variables; the 
Kendall test statistic ($\tau_K$), and corresponding probability of the correlation
arizing by chance ($P_K$); the Spearman's correlation coefficient ($r_{SR}$),
and corresponding probability ($P_{SR}$).

All 15 pairs of luminosities are highly correlated
in the total sample. All of the correlations are characterized by the probability 
$P \leq 10 ^ {-6}$ that the null hypothesis of no correlation is true,
except for the pair $(L_{6cm}, L_H)$, which has a weaker correlation.

However, the results differ when we compare the 3 morphological subsamples:

\noindent
- In the early ($T=0-2$, S0/a-Sab) sample,
the correlations among  $L_{12} , L_{FIR}, L_{6 cm}$ are all
very significant ($P \leq 10 ^ {-6}$). Similarly strong are the correlations of 
$L_B$ with $L_X$ and $L_H$, while the $(L_X, L_H)$ one is marginal.
Typically, correlations among one of $L_{12} , L_{FIR}, L_{6 cm}$ with either
$L_B, L_X, L_H$ are poor or absent.

\noindent
- In the intermediate ($T = 3 - 4$, Sb-Sbc) sample,
strong correlations persist among $L_{12} , L_{FIR}$, and $L_{6 cm}$ 
and between $L_B$ and $L_H$;
$L_X$ is more strongly correlated with the IR
than with either $L_B$ or $L_H$. $L_H$ is now significantly correlated with 
both $L_{12}$ and $L_{FIR}$.

\noindent
- In the late ($T = 5 -10$, Sc-Irr) sample,
all the pairs of variables are very strongly correlated, with $P \leq 10 ^ {-6}$.

Table 6 displays the results of the bivariate
luminosity-ratio correlation tests for the total, early, intermediate, and late
samples. The format is the same as for Table 5.
As is the case for the luminosity correlations discussed above, we find
morphology related differences.
In the total sample we find that X-ray brighter galaxies (for a 
given optical luminosity) are those brighter in the radio continuum, mid and far-IR, and with
warmer far-IR colors. However, these
color correlations only arise in the intermediate and late samples, and are 
absent in the bulge-dominated early sample. 
As discussed in Paper II, these 
effects may all be related to star-formation activity. They also reflect the 
existence of non-linear power-law relations between the luminosities (see below).

Linear regression analysis was applied to bivariate correlations to estimate
the functional relations between the variables.
ASURV's BIVAR offers three routines for linear regression analysis of
censored data:
EM (estimation-maximization) method, Buckley-James method,
and the Schmitt's binning method (\cite{sch85}). 
The first two methods only handle data sets which possess censoring in the
dependent variable alone. Schmitt's method, however,
addresses the problem of censoring
in both variables. Thus, for
many of the luminosity pairs with censoring in both variables, we were 
able to apply only Schmitt's method to perform regression analysis. 
We note, however, that we found very good agreement among the three regression methods
for the luminosity pairs with censoring such that we were able to apply all three.
Instead of defining one variable as "independent" and the other as "dependent," 
for each luminosity pair, $(X,Y)$, in each morphological subgroup, we obtained the
Schmitt's method regression coefficients (slope, intercept, and the uncertainties
in these quantities) 
for both $(X\vert Y)$ and $(Y\vert X)$.
We then used the bisector of these regressions as our final
estimate of the linear relationship between the variables (\cite{iso90}). Appendix A.2
discusses the derivation of these bisectors. 
We did not apply this same analysis to the luminosity-ratios,
because, while the luminosity-ratio pairs display signals of gross correlation,
there is a lot more scatter present in these correlations than in 
the correlations between luminosities, inducing a large uncertainty into any obtained 
value of regression slope.

The power-law dependencies of the bivariate correlations between each pair of
luminosities are given by the slopes of the regression bisectors which are
tabulated in Table 7, together with an estimate of their uncertainty ($\sigma_S$),
and the intercepts ($Int.$) of the bisectors.  These bisector lines, 
along with the regression lines
are plotted on the scatter diagrams of fig. 6. Inspection
of the regression bisectors reveals, first, that different luminosity
pairs are described by different power-law relationships; second, that
the power-law relationship for a given luminosity pair may be a function
of morphological type.

In the total sample, the regression bisectors for the correlations between X-ray, H, far-IR,
and radio countinuum luminosities are consistent (within $2 \sigma$) with linear relations,
i. e. all these luminosities increase in parallel.  
Other correlations are definitely non-linear. These include among others
the well known $L_B \propto L_H^{0.7}$ relation (\cite{ahm79}),
and the strong linear FIR / radio-continuum correlation (\cite{dic84}; \cite{hel85};
\cite{dej85}).
These results are in agreeement with previous studies of large different
representative samples of spiral galaxies, and reinforce our conclusion
of \S2. that our sample is representative of the spiral galaxy population. 
In disagreement with previous reports, we find $L_X \propto L_B^{1.5}$, 
steeper than the relation reported between these
two quantities in \cite{fgt88}, which however was based on the analysis of a much 
smaller sample of 51 galaxies. We will discuss the implications of this result in
Paper II. We suggest there that different mechanism may be responsible for these
effects in early and late-type spirals: hot halos in bulge-dominated galaxies,
and obscuration effects in disk-dominated galaxies and irregulars.

Table 7 shows some morphology-related changes in the relation slopes. 
The best-defined bisectors are in the late sample, where all of the correlations
are very significant. The results of the regression analysis include regression
bisectors that are consistent with a power-law exponent $\alpha \approx 1$ for 
the following luminosity pairs: $(L_X, L_{FIR})$, $(L_{FIR}, L_{6cm})$, $(L_X, L_{6cm})$,
$(L_{6cm}, L_{12})$, and $(L_X, L_H)$. ($L_{6cm}, L_H$) could also be consistent with a
linear trend, but the error is significantly larger for this correlation.
The other pairs exhibit relationships
with power-laws significantly different from unity. These relationships and their possible
implications are discussed in Paper II. We conclude there that the linear relations are
likely to result from the overall connection of those emission bands to star-formation
related phenomena. The non-linear relations point to other effects, including extinction
and possibly the characteristics of the star formation history.

For certain pairs of luminosities, the distribution of early- and 
intermediate-type galaxies spans a smaller range 
in luminosity (typically restricted to higher luminosities), than does the distribution of
late-type galaxies. To derive correlation and regression coefficients,
we simply used all the available data, for each morphological subsample,
regardless of luminosity range.
This approach leads to the question of whether the differences in regression 
slope which we found for different morphological samples [e.g. in the 
$(L_{FIR}, L_{12})$ correlation] may be only an artifact of the different
ranges in luminosity which the different samples span.
In Paper II we address explicitly this questions in the cases where the
results may be affected, by analyzing the data in restricted luminosity ranges.

\section{Multivariate Analysis}

We applied the Partial Spearman Rank analysis
to all of the groups of three, four, five, and six variables
which can be formed from $L_X$, $L_B$, $L_H$, $L_{12}$, $L_{FIR}$, and $L_{6cm}$.
We also held explicitly fixed the distance (D), to verify that
our results are not affected by a distance bias.

The samples used for the multivariate analysis
are smaller than those used to conduct bivariate correlations and regressions,
because we were restricted to include only those galaxies with data for
all six variables. The results for the six variable tests are given in Table 8. Results for 
smaller groupings of variables are tabulated in Appendix A.3.
Table 8 lists the test pair, the parameters held fixed in the test,
the partial rank coefficient, Student t, and corresponding probability of
chance correlation, for the total sample and each of the 3 morphological 
subsamples. The number of points used in each sample is also given ($N$).
Note that the results for the early subsample can only be considered indicative,
given the small number of points. These conclusions are
supported by the analysis of Paper II, which uses the larger samples available for
more limited groupings of variables.

Fig. 10 shows in a diagrammatic form the results of the Partial Rank analysis
for the total, early, intermediate and late samples. Only
the strongest links ($P < 2\%$) are plotted, with their relative strength indicated by
the number of lines connecting variables. 
Two correlations remain very significant, no matter what combination of other variables we
held fixed:  $L_B - L_H$ and $L_{FIR} - L_{12}$.
$L_B$-$L_H$ links stellar emission processes (\cite{ahm79}), and while the presence of a strong
fundamental correlation is not surprising, it also points to a basic connection 
between the IMF of low mass and intermediate-to-high mass stars (\cite{tfb89}).
The tight 12$\mu$m--FIR correlation is consistent with previous findings pointing to
evidence of similarity in the grain size spectrum and distribution in 
the dense ISM of all spirals (see \cite{hel91}, \cite{kgw92}, and refs. therein). 
In this picture the 12$\mu$m emission would be due to small size grains heated to 
non-equilibrium temperature for short times by the same UV photons field responsible 
for the FIR emission.

We find morphology related differences in the correlations.
In the early sample there is an additional strong link of $L_{6cm}$ with $ L_{12}$. 
The intermediate sample results look similar, although the $L_B- L_H$ link
is by far the strongest. The results change in the late sample: the
$L_B- L_H$  link persists, but otherwise we are in the presence of strong 
connections of both 12$\mu$m and  radio continuum with the FIR, 
again suggesting the dominant
effect of star-formation processes in these galaxies (Paper II).
Inspection of Appendix A.3 (Table 11D) shows that most combinations of variables also
yield a significant X-ray -- FIR link in Sc-Irr galaxies, associating
the X-ray emission with the star forming population and associated processes.
This point will be investigated further in Paper II.

Table 9 compares results for the case where the distance is held fixed 
in the analysis, and where is not considered. 
To explore this point further, we performed the Partial Spearman Rank test 
on each pair of variables (holding only the distance fixed), by using the
same sample sizes used in the bivariate analysis. The results (not shown)
compare well with those of Table 5.

While we have $X$ and $B$ data for all of the
galaxies and far-IR data for 93\% of the sample,
our coverage is much sparser in the $H$ and 6cm bands.
The regression analysis of each luminosity pair was performed for 
galaxy samples with data in both of the variables in the luminosity 
pair, regardless of coverage in the other four variables, for
the purpose of using the largest sample possible for each pair. 
Instead, for the multivariate Spearman partial rank analysis, which requires
data for each galaxy in all six of the variables under consideration,
the sample becomes reduced to those galaxies observed in all of the
parameters: $X,B,H,12\mu m, FIR$, and $6 cm$, numbering 94 galaxies.
To explore the effects of the two different sample selections, we performed a 
partial rank analysis on subsamples of variables, by using the largest number of objects 
possible in each case. After checking against the results for the 94 galaxies (6-variable)
sample, we find that our conclusions are generally not affected: while some
correlations are more significant in the larger samples, the relative
strenghts of the different correlations -- which is what we want to 
establish with the multivariate analysis -- follow similar patterns.

\section{Summary and Conclusions}

We have performed bivariate and multivariate survival analyses, which keep
into account censoring (limits),  on a sample of 234 galaxies, covering
morphological types from S0/a to Irregular.  These galaxies were all observed
in X-rays with the {\it Einstein Observatory} (FKT) and their X-ray emission
is not likely to be dominated by an AGN, although some of them may harbor a 
faint active nucleus (i.e. they are representative of normal galaxies in X-rays). 
Besides the X-ray emission, included in the 
analysis were optical (B), near-IR (H), mid- and far-IR, and radio continuum
emissions. Their morphological type was considered explicitly in the 
analysis by dividing the sample in `early' (S0/a-Sab, bulge dominated),
`intermediate' (Sb-Sbc), and `late' (Sc-Irr) subsamples.

In this paper, we have described the sample and the derivation of the variables
used in the analysis; we have reported in details the results of the statistical analysis;
and we have discussed possible biases, to conclude that our overall results are not likely 
to be affected in any major way, by either distance bias, incomplete data coverage,
and beam-size effects.

We find that most pairs of luminosities are correlated when considered individually.
A regression analysis demonstrates that different correlations follow  
different power-law relations. Some of these power-laws are morphology 
dependent. These effects and their significance are discussed further  
in Paper II.

When we ask which of these correlations are likely to be 
fundamental, and which instead may arise from secondary effects, we find
that only two are consistently very strong, regardless of galaxy morphology. 
These are the $L_B - L_H$ and the $L_{12} - L_{FIR}$ correlations.
The former links stellar emission processes {\cite{ahm79}), and points to a basic
connection between the IMF of low-mass and intermediate-to-high-mass
stars (e.g. \cite{tfb89}). The latter may be related to the heating of small and larger
size dust grains by the same UV photon field (e.g. \cite{hel91}).

Other highly significant `fundamental' correlations exist, but are
morphology-dependent. In particular, in S0/a-Sab (and also, but possibly
less strikingly in Sb-Sbc) galaxies we observe
a strong link of radio-continuum and 12$\mu$m (not FIR) emission,
while in Sc-Irr, the strong link is with FIR (not 12$\mu$m) emission.
These differences we will explore further in Paper II.

We also find that in the late sample (Sc-Irr) there is an indication
of an overall connection of X-ray, mid and far-IR, and radio-continuum
emission, which could be related to the presence of star-forming activity
in these galaxies (see also Paper II).

\acknowledgments

We thank Louis Ho, Jonathan Mc Dowell, and Kim McLeod for their useful
input and interest in this work, which was part of the Harvard undergraduate senior thesis 
of A. Shapley. We thank John Huchra for discussions on galaxy distances.
We used the NASA Extragalactic Database (NED) to gather some 
of the data used in this paper. The ADS abstract service was of help in our literature
search. This work was supported by NASA grant NAGW--2681 (LTSA), and by NASA contract 
NAS~8--39073 (Chandra X-ray Center).

\newpage
\appendix
\section{1. Distances}

For this paper we have revised the distances used in the FKT catalog.
The motivation was that recent accurate direct measurements from local 
indicators exist for nearby galaxies, which make up a large fraction of the
sample.
We have performed a thorough literature search through November 1999
to determine the most reliable, up to date distances for our sample. If a recent and
reliable distance estimate was not found, we adopted $H_0$ distances
for $H_0$=75 ${\rm km~s^{-1}~Mpc^{-1}}$, derived from the Yahil, Tammann
\& Sandage (1977; YTS hereafter) corrected velocity.
Heliocentric velocities ($V_0$) were taken from NED.
For each galaxy, Table 10 lists the adopted modulus and distance, 
followed by the heliocentric velocity, the YTS
corrected velocity, and $H_0$ distances for 
$H_0$=75 ${\rm km~s^{-1}~Mpc^{-1}}$. 
For many galaxies a modulus and distance are not listed in columns 2 and 3.  
This is because there are no modern distance estimates available in the 
literature.  In those cases where we give no $H_0$ distance, the actual
measured distance is solid enough that there is no defensible reason for not
using it.  

To estimate the uncertainties that may arise from applying different corrections 
to the heliocentric velocities, 
we also estimated velocities relative to the Cosmic Microwave Background (CMB) frame, 
using a code provided by John Huchra (private communication).
The plot of the fractional difference between YTS and CMB velocities (fig.~11)
for galaxies with Hubble flow distances
shows that differences are within 20\% for $V > 1500$km/s and within 30\%
down to 1000 km/s. Seven more nearby galaxies have differences between 40\% and
60\%. In \S5. we discuss how these uncertainties do not produce significant
differences in the results of our correlation analysis.

Below, we give detailed notes and references.

\noindent
\underline {Notes on groups}:

\noindent
\underline {The Local Group}:
We adopt a distance modulus of 18.50 for the LMC (Madore \& Freedman 1998).  
While there are competing, generally shorter, distance moduli for the LMC in 
the recent literature (e.g.~Luri \etal~1998), the range of values under 
discussion is small:  a systematic uncertainty of $\sim$0.2 magnitudes in the 
zero-point of the distance modulus will not effect the results of this study.  
A distance modulus of 18.50 gives a physical distance of 50 kpc.  The SMC has a 
distance modulus greater than that of the LMC by $\sim$0.4 magnitudes.  
Although studies differ on the zero-point of the distance scale, nearly all of 
them are consistent with this difference in the distances to the two Clouds 
(e.g.~B\"ohm-Vitense 1997).  We thus adopt an SMC distance modulus of 18.90, 
corresponding to a distance of 60 kpc.  For our other Local Group objects, we
adopt Cepheid distances tied to the adopted modulus for the LMC.  For NGC 224 
(M31), IC 1613, and NGC 598 (M33) we adopt the result of Freedman \& Madore 
(1991).  For NGC 6822, we adopt the result of Gallart, Aparicio \& V\'\i lchez 
(1996). 

\noindent
\underline {The Sculptor Group}:
We adopt the Cepheid distance to NGC 300 from Freedman \etal~(1992).  We note 
that there is evidence of a substantial distance spread amongst Sculptor group 
members (Puche \& Carignan 1988).  We thus adopt the relative distances from 
Puche \& Carignan (1988) between NGC 300 and our sample:  $\Delta(m-M)$ = 0.74 
for NGC 247; $\Delta(m-M)$ = 0.79 for NGC 253; $\Delta(m-M)$ = 1.37 for NGC 
7793.  C\^ot\'e \etal~(1997) argue that NGC 625 is a Sculptor group member, 
lying between the main concentration, and NGC 45.  We adopt a distance of 4.9 
Mpc based on the relative velocities and distances of NGC 7793, NGC 45, and NGC 
625.

\noindent
\underline {The IC 342/Maffei 1 group}:
Krismer, Tully \& Gioia (1995) derive Tully-Fisher distances to NGC 1560 and
UGCA 105.  The mean of these measures gives a group distance of 3.6 Mpc.  The 
best distance estimate for NGC 1569 is that of Karachentsev \etal~(1997), who 
derive a distance of 1.7 Mpc from bright stars.  Krismer \etal~(1995) find that 
NGC 1569 does not yield a plausible Tully-Fisher distance.  

\noindent
\underline{NGC 1533 \& NGC 1566 (The Dorado Group)}:  The mean velocity of 11 group members
tabulated in Ferguson \& Sandage (1990) is 1342 km/sec.  We adopt this
for both galaxies, and compute and $H_0$ distance.

\noindent
\underline{NGC 2775 \& NGC 2777}:  We use an $H_0$ distance, based on the mean velocity of the
two group members.

\noindent
\underline{NGC 2992 \& NGC 2993}:  We use an $H_0$ distance, based on the mean velocity of the
two group members.

\noindent
\underline {The M81 group}:
We adopt the Freedman \etal~(1994) Cepheid distance to M81, and use
this distance for NGC 3034, NGC 3077, IC 2574, and NGC 4236.  For NGC 2366, we
adopt the Cepheid distance from Tolstoy \etal~(1995).  For NGC 2403, we adopt
the Cepheid distance from Freedman \& Madore (1988).

\noindent
\underline {The Leo I group}:
We adopt the Cepheid distance to M96 (NGC 3368) and NGC 3489 from Kennicut et al (1998).

\noindent
\underline {The CVn I cloud}:
We adopt recent bright-star distances for the following members of our sample:  
NGC 4190 from Tikhonov \& Karachentsev (1998); NGC 4214 from Makarova, 
Karachentsev \& Georgiev (1997); NGC 4244 from Karachentsev \& Drozdovksy 
(1998). 

\noindent
\underline {The M101 group}:
Stetson \etal (1998) state:  "An unweighted average of 
the two [Cepheid-based] moduli is 29.28 +/- 0.14 mag (with the uncertainty of 
the LMC modulus having been subtracted from the uncertainty each of the two 
estimates and added back in to the uncertainty of the average), implying a 
distance of 7.2 +/- 0.5 Mpc."  This is for M101 (NGC 5457).  We adopt this 
result for NGC 5204, NGC 5474, NGC 5477, and NGC 5585 also.

\noindent
\underline {The Cen A group}:
Saha \etal~(1995) derive a Cepheid distance for NGC 5253.  For NGC 5236 
(M83) Eastman, Schmidt \& Kirshner (1996) derive a SNII expanding photosphere 
distance.

\noindent
\underline {The NGC 3166 group}:
Garcia \etal~(1996) derive a group-average Tully-Fisher distance of 8.8 Mpc.  
We adopt this for both NGC 3166 and NGC 3169.

\noindent
\underline {The Ursa Minor Cluster}:
Pierce \& Tully (1988) derive a mean Tully-Fisher distance of 15.5 Mpc for the Ursa 
Minor cluster.  We adopt this distance for NGC 3729, NGC 3893, NGC 3896, NGC 
4051, IC 749 and IC 750.

\noindent
\underline {The Fornax Cluster}:
Shanks (1997) quotes a Cepheid distance for NGC 1365 of 18.4.  We adopt this 
distance for NGC 1317, NGC 1350, and NGC 1386 as well.  For NGC 1380, we adopt 
the surface brightness fluctuation (SBF) distance reported by Hamuy 
\etal~(1996). 

\noindent
\underline {The Virgo Cluster}:
Given the evidence for substantial depth to the Virgo cluster (e.g.~Yasuda, 
Fukugita \& Okamura 1997), we adopt individual distance estimates to Virgo 
members as follows:  For NGC 4321 (M100), we adopt the Cepheid distance from
Freedman \etal~(1994).  For NGC 4536 we adopt the Cepheid distance from Saha 
\etal~(1996).  For NGC 4571, we adopt the bright-star distance of Pierce, 
McClure \& Racine (1992).  For NGC 4579, we adopt the SNII expanding 
photosphere distance from Eastman \etal~(1996).  For NGC 4639, we adopt the 
Cepheid distance from Sandage \etal~(1996).  Sch\"oniger \& Sofue (1997) derive 
distances for NGC 4303, NGC 4438, and NGC 4647, based on combined CO and H\,I 
Tully-Fisher.  For NGC 4429, we adopt the fundamental plane distance of
Gavazzi et al 1999.
For NGC 4527 we adopt the SNIa distance from 
Shanks (1997).  Teerikorpi \etal~(1992) give Tully-Fisher distances for NGC 
4567 and NGC 4845.  Yasuda \etal~(1997) give $B$-band Tully-Fisher distances 
for a large sample of Virgo galaxies.  The Yasuda \etal~(1997) distances match 
the available Cepheid distances within the errors.  We adopt the Yasuda \etal 
(1997) distances for the following galaxies:  NGC 4178, NGC 4192, NGC 4206, NGC 
4212, NGC 4216, NGC 4235, NGC 4254, NGC 4298, NGC 4351, NGC 4388, NGC 4394, NGC 
4424, NGC 4450, NGC 4501, NGC 4522, NGC 4535, NGC 4548, NGC 4569, NGC 4651, NGC 
4654, NGC 4689, NGC 4698.  We adopt the Yasuda \etal~(1997) mean Virgo distance 
of 16.0 Mpc for NGC 4643, and NGC 4665.
We adopt the Gavazzi \etal (1999) distances for NGC 4461, NGC 4464, NGC 4477, NGC 4503.

\noindent
\underline {Systems behind the Virgo cluster}:
Yasuda \etal~(1997) also report Tully-Fisher distances for the following 
galaxies in the background of the Virgo cluster:  NGC 4224, NGC 4246, NGC 4260, 
NGC 4378.

\noindent
\underline {The Grus Group}:
We adopt a mean $H_0$ distance for the group members (NGC 7496, NGC 7552, NGC 7582,
NGC 7590, NGC 7599).

\noindent
\underline {Notes on individual galaxies}:

\noindent
NGC 628 (M74):  Sharina, Karachentsev \& Tikhonov (1996) derive a distance 
based on bright stars.  They find similar distances for several of M74s dwarf 
companions.  Their result is roughly between the very discrepant results from 
older studies. Distance confirmed by Sohn \& Davidge (1996).

\noindent
NGC 672:  We adopt the result of Sohn \& Davidge (1996), who derive a distance 
for NGC 672 based on bright stars.

\noindent
NGC 1313:  Ryder \etal~(1995) cite a mean distance of 4.5 Mpc based on tertiary 
distance estimators.  They further state that there is no discrepancy between
the long- and short-scale distance camps in the cited work.

\noindent
NGC 1559:  We adopt the SNII expanding photosphere distance from Eastman 
\etal~(1996).   

\noindent
NGC 2441:  We adopt the SNIa distance from Riess, Press \& Kirshner (1996).

\noindent
NGC 3351:  Graham et al. 1997 quote a Cepheid-based distance 
modulus of $30.01 +/- 0.19$, corresponding to a distance of $10.05 +/- 0.88$ Mpc.

\noindent
NGC 3368:  Kennicutt et al. 1998, ApJ 498 181 quote a Cepheid-based distance 
modulus of $30.27 +/- 0.13$, corresponding to a distance of 11.3 Mpc.

\noindent
NGC 3628:  There are no direct distance estimates.  NGC 3628 is a member of the 
NGC 3627 group (Garcia 1993).  Theureau \etal~(1997) quote a Cepheid-based 
distance to NGC 3627, and we adopt this distance for NGC 3628.

\noindent
NGC 4258:  (M106) Herrnstein \etal (1999) 
derive a geometric distance of 7.2 Mpc for a distance modulus of
$(m-M)_0 = 29.29$.

\noindent
NGC 4449:  We adopt the bright-star distance from Karachentsev \& Drozdovsky 
(1998). 

\noindent
NGC 4565:  We adopt the result of Forbes (1996), which is based on an average 
of the results from the globular cluster luminosity function, SBF, and the 
planetary nebula luminosity function (PNLF).

\noindent
NGC 4594 (M104):  We adopt the SBF distance from Ajhar \etal~(1997).

\noindent
IC 4182:  We adopt the Cepheid distance from Saha \etal~(1994).

\noindent
NGC 5037:  There are no direct distance estimates.  However Ferguson \&
Sandage (1990) list NGC 5037 as a member of the NGC 5044 group.  deVaucouleurs 
\& Olson (1984) give Faber-Jackson distances for two group members (NGC
5017 and NGC 5044).  Tutui \& Sofue give a distance for NGC 5054 based on the
average of H\,I and CO Tully-Fisher.  We adopt the mean of these distances for
NGC 5037.

\noindent
NGC 5194 (M51):   We adopt the PNLF distance from Feldmeier, Ciardullo \&
Jacoby (1997).

\noindent
NGC 6503:  We adopt the bright-stars distance from Karachentsev \& Sharina 
(1997).

\noindent
NGC 6946:  Pierce (1994) gives a Tully-Fisher distance of 5.5 Mpc.  Schmidt 
\etal~(1994) give an SNII expanding photosphere distance of 5.7 Mpc.  
Sch\"oniger \& Sofue (1994) give a CO Tully-Fisher distance of 5.4 Mpc.  We 
adopt 5.5 Mpc.

\noindent
NGC 7331:  We adopt the Cepheid distance from Hughes \etal~(1998). 

\noindent
Tutui \& Sofue (1997) derive distances based on the average of CO and H\,I 
Tully-Fisher that we adopt for the following members of our sample:  NGC 520,
NGC 772, NGC 1961, NGC 4038.

\noindent
Sch\"oniger \& Sofue (1994) derive distances from the average of H\,I and CO 
Tully-Fisher, that we adopt for the following members of our sample:  NGC 2276, 
NGC 3079, NGC 4631, NGC 4736, NGC 5907, NGC 7469.  IC 5283 is a companion of 
NGC 7469, and we adopt the same distance as NGC 7469.

\noindent
We adopt distances quoted by Shanks (1997), based on Cepheids or SNIa, for the 
following members of our sample:  NGC 2841, NGC 3351, NGC 3389.
The distance to NGC 3351 is confimed by Graham \etal (1997).

\noindent
We adopt SNIa and SNII distances from Pierce (1994) for the following members 
of our sample:  NGC 3184, NGC 7339.

\newpage

\appendix
\section{2. Calculation of the Regression Bisectors}

The bisector slope $(\beta _{bis})$ and intercept($\alpha_{bis}$), where estimated using
the following expressions from Isobe et al (1990):
\begin{equation}
\beta _{bis} = (\beta _1 + \beta _2)^{-1} [\beta_ 1 \beta _2 - 1 + 
 \sqrt{(1 + \beta _1 ^ 2)(1 + \beta _2 ^ 2 )}]
\end{equation}

\begin{equation}
\alpha _{bis} = y_{int} -  \beta _{bis}x_{int} 
\end{equation}

\noindent where $y_{int}$ and $x_{int}$ are the coordinates of the intersection
point of two the regressions, $y = \beta _1x + \alpha_1$ and $ y = \beta _2x + \alpha_2$.

The bisector slope, $\beta _bis$ is a function of two interdependent variables, $\beta _1$ and
 $\beta _2$.
Therefore, in order to find the uncertainty in $\beta _bis$, we need to calculate 
the following (from \cite{iso90}):

\begin{equation}
\sigma_{\beta_ {bis}}^2 = \sigma_{\beta_ {1}}^2 (\frac {\partial \beta _{bis}}{\partial \beta 
_1} )^2 
+ \sigma_{\beta_ {2}}^2(\frac {\partial \beta _{bis}}{\partial \beta _2} )^2 +
2\sigma _{\beta _1 \beta _2} (\frac {\partial \beta _{bis}}{\partial \beta _1} )
(\frac {\partial \beta _{bis}}{\partial \beta _2} )
\end{equation}

\noindent where $\sigma_{\beta_ {1}}$ and $\sigma_{\beta_ {2}}$ are the uncertainties on the slopes, 
$\beta _1$ and $\beta _2$, 
respectively; $\frac {\partial \beta _{bis}}{\partial \beta _1}$ and
$\frac {\partial \beta _{bis}}{\partial \beta _2}$ are the respective partial
derivatives of $\beta _{bis}$ with respect to $\beta _1$ and $\beta _2$;
$\sigma _{\beta _1 \beta _2}$ is the covariance of $\beta _1$ and $\beta _2$.

We obtained $\sigma_{\beta_ {1}}$ and $\sigma_{\beta_ {2}}$ from the Schmitt's
regression analysis package in ASURV, which provides a bootstrap error
analysis. The expressions for the partial derivatives are:

\begin{equation}
\frac {\partial \beta _{bis}} {\partial \beta _1} = \frac {(1 + (\beta _2)^2) \beta _{bis}} 
{(\beta _1 + \beta _2)\sqrt { (1 + \beta _1 ^ 2)(1 + \beta _2 ^ 2 ) } }
\end{equation}

\begin{equation}
\frac {\partial \beta _{bis}}{\partial \beta _2} = \frac {(1 + (\beta _1)^2) \beta _{bis}}
{(\beta _1 + \beta _2)\sqrt{(1 + \beta _1 ^ 2)(1 + \beta _2 ^ 2 )} } 
\end{equation}

According to \cite{iso90}, $\sigma _{\beta _1 \beta _2}$, the covariance term,
is calculated in the following manner:

\begin{equation}
\sigma _{\beta _1 \beta _2} = \frac {\beta _1}{ (\sum_{i = 1}^{n} (x_i - \overline{x}) ^2) ^2}
\sum_{i = 1}^{n} (x_i - \overline{x})(y_i - \overline{y})
[(y_i - \overline{y}) - \beta _1 (x_i - \overline{x})]
[(y_i - \overline{y}) - \beta _2 (x_i - \overline{x})]
\end{equation}

\noindent where $\overline{x}$ and $\overline{y}$ are the sample means and $n$ is the number
of data points.

Since the covariance depends explicitly on the 
coordinates of the data points in the sample, it is not obvious how to calculate it in
the presence of censoring.  We estimated the magnitude
of the covariance term for all of the samples of data points, treating the upper limits
as detections, in order to see if this term could be neglected in the calculation
of the bisector slope uncertainty. We found that typically the covariance term
is much smaller than the other two terms which are included in the expression for
the bisector slope uncertainty. Thus, we decided to approximate the bisector slope
uncertainty as:

\begin{equation}
\sigma_{\beta_ {bis}}^2 \approx  \sigma_{\beta_ {1}}^2 (\frac {\partial \beta _{bis}}{\partial
 \beta _1} )^2
+ \sigma_{\beta_ {2}}^2 (\frac {\partial \beta _{bis}}{\partial \beta _2} )^2 
\end{equation}

The plots themselves (fig. 4) offer a visual representation of the uncertainty
of each bisector slope, which depends on the strength of the correlation between
the two variables. The stronger the correlation, the smaller the angle between
the two regressions, and the better-defined the bisector slope.

\newpage
\appendix
\section{3. Results of Spearman Partial Rank Tests}
Tables 11A, B, C, and D list the results of the Partial Rank analysis applied to
each pair of variables for the total sample and the three subsamples.

\clearpage




\clearpage

%
%
%


\clearpage
\figcaption{Distribution of absolute magnitudes for the sample galaxies. }

\figcaption{Distribution of sample galaxies in morphological types (T). The unshaded
regions denote the galaxies flagged as AGN. }

\figcaption{Comparison of our calculated $H_{0.5}$ with those of Tormen \& Burstein (1995).}

\figcaption{Distributions of X-ray luminosities in the Total sample and the three 
morphological subsamples (`Early', T=0-2; `Intermediate', T=3-4; and `Late', T=5-10).
For comparison we also show the distribution of $L_X$ for the FKT E and S0 galaxies.
In all diagrams, except the T=0-2 one, the shaded area represents detections, 
the unshaded area represents upper limits. In the T=0-2 diagram different levels of
shading represent: unshaded -- S0/a-Sab upper limits; light shading -- S0/a-Sab detections;
heavier shading -- Amorphous upper limits; solid shading -- Amorphous detections.
}

\figcaption{Distributions of $L_X /L_B$ in the Total sample and the three 
morphological subsamples (`Early', T=0-2; `Intermediate', T=3-4; and `Late', T=5-10).
For comparison we also show the distribution of $L_X /L_B$ for the  FKT E and S0 galaxies.
Same shading conventions as in fig.~2.
}

\figcaption{Scatter diagrams for luminosity pairs. For each pair, the scatter diagrams
for the Total sample and for the three morphological subsamples (`Early', T=0-2; 
`Intermediate', T=3-4; and `Late', T=5-10) are plotted. Filled squares identify detections
on both axes; triangles identify upper limits in one of the axis, with the apex pointing
in the direction of the limit; empty circles identify upper limits in both axes;  circles
surrounding another symbol identify the flagged AGN, which were not included
in the statistical analysis; squares surrounding another symbol identify Amorphous galaxies, 
which were not included in the T=0-2 analysis. 
The solid lines across the points represent the regression
bisectors, while individual regressions are represented by the two dashed lines. }

\figcaption{Scatter diagrams of Log($L_X/L_B$) versus other luminosity ratios.
For each pair, the scatter diagrams
for the Total sample and for the three morphological subsamples (`Early', T=0-2;
`Intermediate', T=3-4; and `Late', T=5-10) are plotted. Filled squares identify detections
on both axes; triangles identify upper limits in one of the axis, with the apex pointing
in the direction of the limit; empty circles identify upper limits in both axes;  circles
surrounding another symbol identify the flagged AGN, which were not included
in the statistical analysis; squares surrounding another symbol identify Amorphous galaxies, 
which were not included in the T=0-2 analysis. }

\figcaption{Log($L_X$) -- Log(D) scatter diagram for the Total sample. Circles are AGN
detections, squares are detections, and triangles are upper limits.}

\figcaption{Log($L_X$) -- Log($L_{6cm}$) scatter diagrams for the Total (a) and Late T=5-10
sample (b). Different symbols are used for different galaxy diameters; see fig. 9a. We
do not detect any evident displacement of large diameter galaxies. }

\figcaption{Graphical representation of the Partial Spearman Rank analysis. Significant 
correlations are represented by lines connecting the variables, with a greater number 
of connecting lines identifying relatively stronger correlations. Detailed test results 
are given in Table 8. We show diagrams for the three morphological subsamples (`Early', T=0-2;
`Intermediate', T=3-4; and `Late', T=5-10) and for the total sample. }

\figcaption{Fractional difference between YTS and CMB corrected Hubble flow velocities 
versus the YTS velocity}


\begin{thebibliography}{}


\bibitem[Aaronson, Huchra \& Mould 1979]{ahm79}Aaronson M., Huchra, J., 
Mould, J. 1979, \apj, 229, 1. 



\bibitem[Aaronson \etal 1981]{aaretal81}Aaronson, M., Dawe, J.A., 
Dickins, R.J., Mould, J.R., Murray, J.B. 
1981, \mnras, 195, 1P

\bibitem[Aaronson \etal 1982]{ahm82}
Aaronson M., Huchra, J., Mould, J., Tully, R.B., Fisher, J.B.,
Van Woerden, H., Goss, W.M., Chamaraux, P., Mebold, U.,
Siegman, B., Berriman, G., Persson, S.E. 1982, \apjs, 50, 241.

\bibitem[]{}Ajhar, E.A., Lauer, T.R., Tonry, J.L., Blakeslee, J.P., Dressler, A., Holtzman, 
J.A. \& Postman, M. 1997, \aj, 114, 626


\bibitem[Allen 1973]{all73}Allen, C.W. 1973, Astrophysical Quantities (London: The Athlone Press), p.197.

\bibitem[]{}Allen, D.A. 1976, ApJ, 207, 367

\bibitem[Balzano \& Weedman 1981]{baw81}Balzano, V.A., Weedman, D.W. 1981, \apj, 243, 756

\bibitem[Becker, White \& Edwards 1991]{bec91}Becker, R.H., White, R.L., Edwards, A. 1991, \apjs, 75,1

\bibitem[Becklin \etal 1980]{bgm1980}Becklin, E. E., Gatley, I., Matthews, K., Neugebauer, G.,
Sellgren, K., Werner, M. W., Wynn-Williams, C. G.  1980,
\apj, 236, 441

\bibitem[Beichman \& Neugebauer 1984]{bei84}Beichman, C. A., Neugebauer, G. 1984, Infrared
Astronomical Satellite (IRAS) Catalogs and Atlases Explanatory Supplement (Pasadena:JPL)

\bibitem[]{}B\"ohm-Vitense, E. 1997, \aj, 113, 13

\bibitem[Bothun \etal 1984]{bas84}Bothun, G.D., Aaronson, M., Schommer, B., Huchra, J., Mould, J. 1984,
\apj, 278, 475

\bibitem[Bothun \etal 1985]{bas85}Bothun, G.D., Aaronson, M., Schommer, B., Mould, J. 
Huchra, J., Sullivan, W.T. III 1985, \apjs, 57, 423

%
%
%
%
%
%


\bibitem[Colbert \& Mushotzky 1999]{cm99}
Colbert, E. J. M. \& Mushotzky, R. F. 1999, \apj, 519, 89


\bibitem[Calvani, Fasano \& Franceschini 1989]{cal89}Calvani, M., 
Fasano, G., Franceschini, A. 1989, \aj, 97, 1319

\bibitem[Condon 1980]{con80}Condon, J.J. 1980, \apj, 242, 894

\bibitem[Condon \etal 1982]{con82}Condon, J.J., Condon, M.A., Gisler, G., Puschell, J.J.
1982, \apj, 252, 102

\bibitem[Condon, Frayer \& Broderick 1991]{con91}Condon, J.J., Frayer, D.T, Broderick, J.J 1991, \aj, 101, 362

\bibitem[Corbelli, Salpeter \& Dickey 1991]{cor91}Corbelli, E., Salpeter, E., Dickey, J. 1991, \apj, 370, 49

\bibitem[]{}C\^ot\'e, S., Freeman, K.C., Carignan, C., \& Quinn, P.J. 1997, \aj, 114, 1313.

\bibitem[Cutri \& McAlary 1985]{cum85}Cutri, R.M., McAlary, C.W. 1985, \apj, 296, 90

\bibitem[de Jong \etal 1985]{dej85}de Jong, T., Klein, U., Wielebinski, R., Wunderlich, E.
1985, \aap, 147, L6

\bibitem[de Vaucouleurs \& Longo 1988]{dev88}de Vaucouleurs, A., Longo, G. 1988, Catalogue of
Visual and Infrared Photometry of Galaxies From 0.5 $\mu$m to 10 $\mu$m
(1961 - 1985) (Austin: University of Texas Press)

\bibitem[]{}deVaucouleurs, G. \& Olson, D.W. 1984, \apjs, 56, 91

\bibitem[de Vaucouleurs, de Vaucouleurs \& Corwin 1976]{RC2}de Vaucouleurs, G., de Vaucouleurs A., Corwin, H.G. 1976,
Second Reference Catalogue of Bright Galaxies (Austin:
University of Texas Press) (RC2)

\bibitem[de Vaucouleurs \etal 1991]{RC3}de Vaucouleurs, G., de Vaucouleurs
 A., Corwin, H.G.jr, Buta, R. J., Paturel, G., Fouque, P. 1991, The Third Reference
Catalogue of Bright Galaxies, (New York: Springer) (RC3)


\bibitem[Dickey \& Salpeter 1984]{dic84}Dickey, J.M., Salpeter, E. E. 1984, \aj, 284, 461

\bibitem[Disney \& Wall 1977]{dis77}Disney, M.J., Wall, J.V. 1977, \mnras, 179, 235

\bibitem[]{}Eastman, R.G., Schmidt, B.P. \& Kirshner, R. 1996, \apj, 466, 911

\bibitem[Ekers \& Ekers 1973]{eke73}Ekers, R.D., Ekers, J.A. 1973, \aap, 24, 247

\bibitem[Elvis, Soltan \& Keel 1984]{esk84}Elvis, M., Soltan, A., Keel, W.C. 1984, \apj, 283, 479


\bibitem[Eskridge, Fabbiano \& Kim 1995a]{efk95}Eskridge, P.B., Fabbiano, G., Kim, D.-W. 1995a, \apjs, 97, 141

\bibitem[Eskridge, Fabbiano \& Kim 1995b]{efk2}Eskridge, P.B., Fabbiano, G., Kim, D.-W. 1995b, \apj, 442, 523

\bibitem[Eskridge, Fabbiano \& Kim 1995c]{efk2}Eskridge, P.B., Fabbiano, G., Kim, D.-W. 1995c, \apj, 448, 70

\bibitem[Fabbiano 1989]{f89}Fabbiano, G. 1989, Ann.Rev.A.Ap., 27, 87.

\bibitem[Fabbiano 1990]{f90}Fabbiano, G. 1990 in {\it Windows on Galaxies}, 
eds. G. Fabbiano,
J.S. Gallagher, A. Renzini, p. 231. Dordrecht: Kluwer

\bibitem[Fabbiano, Gioia \& Trinchieri 1988]{fgt88}Fabbiano, G., Gioia, 
I.M, Trinchieri, G. 1988, \apj, 324, 749

\bibitem[Fabbiano, Gioia \& Trinchieri 1989]{fgt89}
Fabbiano, G., Gioia, I.M., Trinchieri, G. 1989, \apj, 347, 127

\bibitem[Fabbiano, Kim \& Trinchieri 1992]{fkt92}
Fabbiano, G., Kim, D.-W., Trinchieri, G. 1992, \apjs, 80, 531.
(FKT)

\bibitem[Fabbiano \& Shapley 2001]{fs97}Fabbiano, G., Shapley, A. 2001, in preparation (Paper II).

\bibitem[Fabbiano \& Trinchieri 1985]{ft85}Fabbiano, G., Trinchieri, G. 1985, \apj, 296, 430

\bibitem[Fabbiano \& Trinchieri 1987]{ft87}Fabbiano, G., Trinchieri, G. 1987, \apj, 315, 46.

\bibitem[Feigelson \& Berg 1983]{fb83}Feigelson, E.D., Berg, C.J. 1983, \apj, 269, 400

\bibitem[]{}Feldmeier, J.J., Ciardullo, R. \& Jacoby, G.H. 1997, \apj, 479, 231

\bibitem[]{}Ferguson, H.C. \& Sandage, A. 1990, \aj, 100, 1

\bibitem[]{}Forbes, D.A. 1996, \aj, 112, 1409

\bibitem[]{}Freedman, W.L., Hughes, S.M., Madore, B.F., Mould, J.R., Lee, M.G., Stetson, 
P., Kennicutt, R.C., Turner, A., Ferrarese, L, Ford, H., Graham, J.A., Hill, 
R., Hoessel, J.G., Huchra, J., \& Illingworth, G.D. 1994, \apj, 427, 628

\bibitem[]{}Freedman, W.L., \& Madore, B.F. 1988, \apj, 332, L63

\bibitem[]{}Freedman, W.L., \& Madore, B.F. 1991, \pasp, 103, 933

\bibitem[]{}Freedman, W.L., Madore, B.F., Hawley, S.L, Horowitz, I.K., Mould, J., 
Navarrete, M., \& Sallmen, S. 1992, \apj, 396, 80

\bibitem[Frogel \etal 1978]{fpa78}
Frogel, J.A., Persson, S.E., Aaronson, M., Matthews, K. 1978,
\apj, 220, 75.

\bibitem[Fullmer \& Lonsdale 1989]{ful89}
Fullmer, L., Lonsdale, C. 1989, Catalogued Galaxies
and Quasars Observed in the {\it IRAS} Survey, Version 2 
(Pasadena: JPL) 

\bibitem[Gallagher \& Fabbiano 1990]{gf90}Gallagher, J, Fabbiano, G. 1990, 
in {\it Windows on Galaxies}, eds. G. Fabbiano, J.S. Gallagher, A. Renzini, p.1. 
Dordrecht: Kluwer

\bibitem[Gallagher \etal 1991]{gal91}
Gallagher, J.S., Hunter, D.A., Gillett, F.C.,
Rice, W.L. 1991, \apj, 371, 142. 

\bibitem[]{}Gallart, C., Aparicio, A., \& V\'\i lchez, J.M. 1996, \aj, 112, 1928

\bibitem[]{}Garcia, A.M. 1993, \aaps, 100, 47

\bibitem[]{}Garcia, A.M., Fournier, A., DiNella, H. \& Paturel, G. 1996, \aap, 310, 412

\bibitem[Giacconi \etal 1979]{g79}Giacconi, R. \etal 1979, \apj, 230, 540.

\bibitem[Gioia \& Fabbiano 1987]{gf87}
Gioia, I.M., Fabbiano, G. 1987, \apjs, 63, 771.

\bibitem[Gioia, Gregorini \& Klein 1982]{ggk82}Gioia, I.M., Gregorini, L., Klein, U.
1982, Astron. Ap., 116, 164

\bibitem[Glass 1976]{gla76}Glass, I.S. 1976, \mnras, 175, 191.

\bibitem[Glass 1984]{gla84}Glass, I.S. 1984, \mnras, 211, 461.

\bibitem[Glass \& Moorwood 1985]{glm85}Glass, I.S., Moorwood, A.F.M. 1985 \mnras, 214, 429.

\bibitem[Golombek, Miley \& Neugebauer 1988]{gmn88}
Golombek, D., Miley, G.K., Neugebauer, G. 1988, \aj, 95, 26.

\bibitem[]{}Graham, J. A. et al  1997, \apj, 477, 535

\bibitem[Griersmith, Hyland \& Jones 1982]{ghj82}
Griersmith, D., Hyland, A.R., Jones, T.J. 1982, \aj, 87, 1106.

\bibitem[]{}Hamuy, M., Phillips, M.M., Suntzeff, N.B., Schommer, R.A., Maza, J. \& Aviles,
R. 1996, \aj, 112, 2391


\bibitem[Haynes et al 1975]{}
Haynes, R. F., Huchtmeir, W. K. G. Siegman, B. C 1975,
A compendium of radio measurements of bright galaxies, (Melbourne:
Commonwealth Scientific and Industrial Research Organization (CSIRO),
Division of Radiophysics)


\bibitem[Heckman \etal 1983]{hec83}
Heckman, T.M., Lebofsky, M.J., Rieke, G.H., Van Breugel, W. 1983,
\apj, 272, 400.

\bibitem[Helou er al 1988]{}Helou, G., Khan, I. R., Malek, L., \& Boehmer, L. 1988,
\apjs, 68, 151

\bibitem[Helou, Ryter \& Soifer 1991]{hel91}Helou, G., Ryter, C., Soifer, B.T. 1991,
\apj, 376, 505.

\bibitem[Helou, Soifer \& Rowan-Robinson 1985]{hel85}
Helou, G., Soifer, B.T., Rowan-Robinson, M. 1985, \apj, 298, L7. 

\bibitem[]{}Herrnstein, J. R., Moran, J. M., Greenhill, L. J., Diamond, P. J.,
Inoue, M., Nakai, N., Miyoshi, M., Henkel, C., \& Riess, A. 1999, Nature, 400, 539

\bibitem[Ho, Filippenko \& Sargent 1997]{hfs97}
Ho, L. C., Filippenko, A. V., \& Sargent, W. L. W. 1997, \apjs, 112, 315

\bibitem[]{}Hughes, S.M.G., Han, M., Hoessel, J., Freedman, W.L., Kennicutt, R.C., Jr., 
Mould, J.R., Saha, A., Stetson, P.B., Madore, B.F., Silbermann, N.A., Harding,
P., Ferrarese, L., Ford, H., Gibson, B.K., Graham, J.A., Hill, R., Huchra, J.,
Illingworth, G.D., Phelps, R. \& Sakai, S. 1998, \apj, 501, 32

\bibitem[Hummel, van der Hulst \& Dickey 1984]{hum84}Hummel, E., van der Hulst, J.M., Dickey, J.M. 1984, \aap,
134, 207.

\bibitem[Hunter \& Gallagher 1985]{hug85}Hunter, Gallagher, J.S. III 1985, \aj, 90, 1457.


\bibitem[Isobe \etal 1990]{iso90}Isobe, T., Feigelson, E.D., Akritas, M.G., Babu, G.J 1990, \apj,
364, 104. 

\bibitem[Jones, Terzian \& Sramek 1981]{jon81}
Jones, D.L., Terzian, Y., Sramek, R.A. 1981, \apj, 247, L57.

\bibitem[]{}Karachentsev, I.D., Drozdovsky, I., Kajsin, S., Takalo, L.O., Heinamaki, P., \&
Valtonen, M. 1997, \aaps, 124, 559

\bibitem[]{}Karachentsev, I.D. \& Drozdovsky, I.O. 1998, \aaps, 131, 1

\bibitem[]{}Karachentsev, I.D. \& Sharina, M.E. 1997, \aap, 324, 457




\bibitem[Kendall \& Stuart 1976]{ken76}Kendall, M., Stuart, A. 1976, The Advanced Theory
of Statistics, Vol. 2 (New York: Macmillan)

\bibitem[Kennicutt et al 1998]{}Kennicutt, R. C. Jr, et al 1998, \apj, 498, 181

\bibitem[Klein 1986]{kle86}Klein, U. 1986, \aap, 168, 65. 

\bibitem[Knapp, Bies \& Van Gorkom, J.H. 1990]{kna90}
Knapp, G.R., Bies, W.E., Van Gorkom, J.H. 1990, \aj, 99, 476.

\bibitem[Knapp \etal 1989]{kna89}Knapp, G.R., Guhathakurta, P., Kim, D.-W., Jura, M. 1989,
\apjs, 70, 329.

\bibitem[Knapp, Gunn \& Wynn-Williams 1992]{kgw92}Knapp, G.R., Gunn, J.E., Wynn-Williams,
C.G. 1992, \apj, 399, 76


\bibitem[]{}Krismer, M., Tully, R.B., \& Gioia, I.M. 1995, \aj, 110, 1584

\bibitem[LaValley, Isobe \& Feigelson 1992]{lav92}
LaValley, M.P., Isobe, T., Feigelson, E.D. 1992, BAAS, 24, 839.

\bibitem[Lonsdale \& Helou 1985]{lon85}
Lonsdale,  C. J., Helou, G. 1985,
Cataloged Galaxies and Quasars Observed in the IRAS Survey, (Pasadena: JPL).

\bibitem[Lonsdale Persson \& Helou 1987]{lon87}
Lonsdale Persson, C. J., Helou, G. 1987, \apj, 314, 513.

\bibitem[]{}Luri, X., Gomez, A.E., Torra, J., Figueras, F. \& Mennessier, M.O. 1998, \aap, 
335, L81

\bibitem[]{}Madore, B.F., \& Freedman, W.L. 1998 \apj, 492, 110

\bibitem[]{}Makarova, L.N., Karachentsev, I.D. \& Georgiev, Ts.B. 1997, AstL, 23, 378

\bibitem[Magorrian et al 1998]{mag98}
Magorrian, J., et al 1998, \aj, 115, 2285





\bibitem[Mould, Aaronson \& Huchra 1980]{mah80}
Mould, J., Aaronson, M., Huchra, J. 1980, \apj, 238, 458.

\bibitem[Mould 1981]{mou81}Mould, J. 1981, \pasp, 93, 25.

\bibitem[Palumbo \etal 1985]{pal85}Palumbo, G.G.C., Fabbiano, G., Fransson, C.,
Trinchieri, G. 1985, \apj, 298, 259.


\bibitem[Persson, Frogel \& Aaronson 1979]{pfa79}
Persson, S.E., Frogel, J.A., Aaronson, M., 1979, \apjs, 39, 61.

\bibitem[Penston \etal 1974]{pen74}Penston, M.V., Penston, M.J., Selmes, R.A., Becklin, E.E.,
Neugebauer, G. 1974, \mnras, 169, 357.

\bibitem[]{}Pierce, M.J., McClure, R.D. \& Racine, R. 1992, \apj, 393, 523

\bibitem[]{}Pierce, M.J. \& Tully, R.B. 1988, \apj, 330, 579

\bibitem[]{}Pierce, M.J. 1994, \apj, 430, 53

\bibitem[]{}Puche, D., \& Carignan, C. 1988, \aj, 95, 1025

\bibitem[Rice \etal 1988]{ric88}Rice, W., Lonsdale, C.J., Soifer, B.T., Neugebauer, G.,
Kopan, E.L., Lloyd, L.A., de Jong, T., Habing, H.J. 1988, \apjs, 68, 91.

\bibitem[Rieke 1978]{rie78}Rieke, G., H., 1978, \apj, 226, 550.

\bibitem[]{}Riess, A.G., Press, W.H. \& Kirshner, R.P. 1996, \apj, 473, 88

\bibitem[Roberts 1991]{rob91}
Roberts, M.S., Hogg, D.E.,
Bregman, J.N., Forman, W.R., Jones, C. 1991, \apjs, 75, 751. 

\bibitem[]{}Ryder, S.D., Staveley-Smith, L., Malin, D. \& Walsh, W. 1995, \aj, 109, 1592

\bibitem[]{}Saha, A.,  Labhardt, L., Schwengeler, H., Macchetto, F.D., Panagia, N., 
Sandage, A. \& Tammann, G.A. 1994, \apj, 425, 14

\bibitem[]{}Saha, A., Sandage, A., Labhardt, L., Schwengeler, H., Tammann, G.A., Panagia, 
N. \& Macchetto, F.D. 1995, \apj, 438, 8

\bibitem[]{}Saha, A., Sandage, A., Labhardt, L., Tammann, G.A., Macchetto, F.D., \& 
Panagia, N. 1996, \apj, 466, 55

\bibitem[]{}Sandage, A., Saha, A., Tammann, G.A., Labhardt, L., Panagia, N. \& Macchetto, 
F.D. 1996, \apj, 460, L15


\bibitem[Sandage \& Tammann 1987]{sand87} Sandage, A., \& Tammann, G. A.
1987, A Revised Shapley-Ames Catalog of Bright Galaxies (2nd ed.,
Washington, DC: Carnegie Institution of Washington)(RSA)

\bibitem[]{}Schmidt, B.P., Kirshner, R.P., Eastman, R.O., Phillips, M.M., Suntzeff, N.B., 
Hamuy, M., Maza, J. \& Aviles, R. 1994, \apj, 432, 42

\bibitem[Schmitt 1985]{sch85}Schmitt, J.H.M.M. 1985, \apj, 293, 178.

\bibitem[]{}Sch\"oniger, F. \& Sofue, Y. 1994, \aap, 283, 21

\bibitem[]{}Sch\"oniger, F. \& Sofue, Y. 1997, \aap, 323, 14

\bibitem[]{}Shanks, T. 1997, \mnras, 290, L77

\bibitem[]{}Sharina, M.E., Karachentsev, I.D. \& Tikhonov, N.A. 1996, \aaps, 119, 499

\bibitem[]{}Sohn, Y.-J. \& Davidge, T.J. 1996, \aj, 112, 25

\bibitem[Sramek 1975]{sra75}Sramek, R. 1975, \aj, 80, 771.

\bibitem[]{}Stetson, P. B. et al 1998, \apj, 508, 491

\bibitem[Stocke, Tifft \& Kaftan-Kassim  1978]{sto78}
Stocke, J.T., Tifft, W.G., Kaftan-Kassim, M.A 1978, \aj, 83, 322.

\bibitem[Sulentic 1976]{sul76}Sulentic, J.W. 1976, \apjs, 32, 171.


\bibitem[]{}Teerikorpi, P., Bottinelli, L., Gouguenheim, L. \& Paturel, G. 1992, \aap, 260,


\bibitem[Telesco \& Harper]{tel80}
Telesco, C.M., Harper, D.A. 1980, \apj, 235, 392

\bibitem[]{}Theureau, G., Hanski, M., Ekholm, T., Bottinelli, L., Gouguenheim, L., Paturel,
G. \& Teerikorpi, P. 1997, \aap, 322, 730

\bibitem[Thuan 1983]{thu83}Thuan, T.X. 1983, \apj, 268, 667.

\bibitem[]{}Tikhonov, N.A. \& Karachentsev, I.D. 1998, \aaps, 128, 325

\bibitem[]{}Tolstoy, E., Saha, A., Hoessel, J.G., \& McQuade, K. 1995, \aj, 110, 1640

\bibitem[Tormen \& Burstein 1995]{tb95}Tormen, G., \& Burstein, D. 1995, \apjs, 96, 123

\bibitem[Trinchieri \& Fabbiano 1991]{tf91}Trinchieri, G., Fabbiano, G. 1991, \apj, 382, 82.

\bibitem[Trinchieri, Fabbiano \& Bandiera, 1989]{tfb89}
Trinchieri, G., Fabbiano, G., Bandiera, R. 1989, \apj, 342, 759
(TFB).

\bibitem[Trinchieri, Fabbiano \& Peres 1988]{tfp88}
Trinchieri, G., Fabbiano, G., Peres, G. 1988, \apj, 325, 531.

\bibitem[Trinchieri, Fabbiano \& Romaine 1990]{tfr90}
Trinchieri, G., Fabbiano, G., Romaine, S. 1990, \apj, 356, 110.

\bibitem[Tully, Mould \& Aaronson 1982]{tma82}
Tully, B.R, Mould, J. R. \& Aaronson, M. 1982, \apj, 2557, 527.

\bibitem[Tully 1988]{tul88}Tully, B.R. 1988, Nearby Galaxies Catalog (New York: 
Cambridge University Press)

\bibitem[]{}Tutui, Y. \& Sofue, Y. 1997, \aap, 326, 915

\bibitem[Ulvestad, Wilson \& Sramek 1981]{ulv81}Ulvestad, J.S., Wilson, A.S., Sramek, R.A. 1981, \apj, 247, 419. 

\bibitem[Ulvestad \& Wilson 1984]{ulv84}Ulvestad, J.S., Wilson, A.S., 1984, \apj, 285, 439.

\bibitem[Ulvestad \& Wilson 1989]{ulv89}Ulvestad, J.S., Wilson, A.S., 1989, \apj, 343, 659.

\bibitem[Ward \etal 1982]{waw82}Ward, M., Allen, D.A., Wilson, A.S., Smith, M.G., Wright, A.E. 1982,
\mnras, 199, 953.

\bibitem[Whiteoak 1970]{whi70}Whiteoak, J.B. 1970, \aplett, 5, 29.

\bibitem[Whitmore 1984]{whi84}Whitmore, B.C. 1984, \apj, 278, 61.

\bibitem[Willick et al 1996]{wil96}
Willick, J. A., Courteau, S., Faber, S. M., Burstein, D., Dekel, A., Kolatt, S. 1996, \apj, 457, 460.

\bibitem[Willner \etal 1985]{wef85}Willner, S.P., Elvis, M., Fabbiano, G., Lawrence, A.,
Ward, M. J. 1985, \apj, 299, 443.

\bibitem[Wilson \& Ulvestad 1982]{wil82}Wilson, A.S., Ulvestad, J.S. 1982, \apj, 260, 56.

\bibitem[Wright 1974]{wri74}Wright, A.E. 1974, \mnras, 167, 273.

\bibitem[Wunderlich, Wielebinski \& Klein 1987]{wun87}
Wunderlich, E., Wielebinski, R., Klein, U. 1987, A \& AS, 69, 487.

\bibitem[Wunderlich \& Klein 1991]{wun91}Wunderlich, E., Klein, U. 1991, A \& AS, 87, 247.

\bibitem[]{}Yahil, A., Tammann, G.A., \& Sandage, A. 1977, \apj, 217, 903. 

\bibitem[]{}Yasuda, N., Fukugita, M. \& Okamura, S. 1997, \apjs, 108, 417

\bibitem[Young 1990]{y90}Young, J.S. in {\it Windows on Galaxies}, eds. G. Fabbiano, 
J.S. Gallagher, A. Renzini, p. 213, Dordrecht: Kluwer. 

\end{thebibliography}
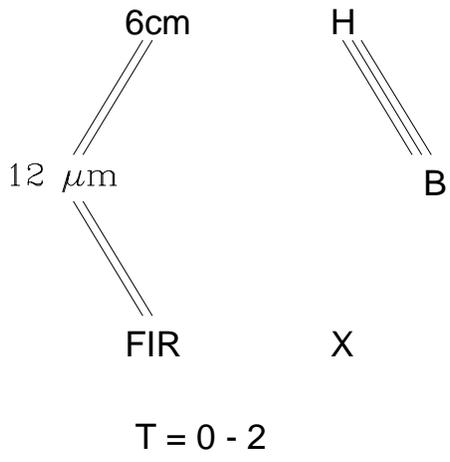
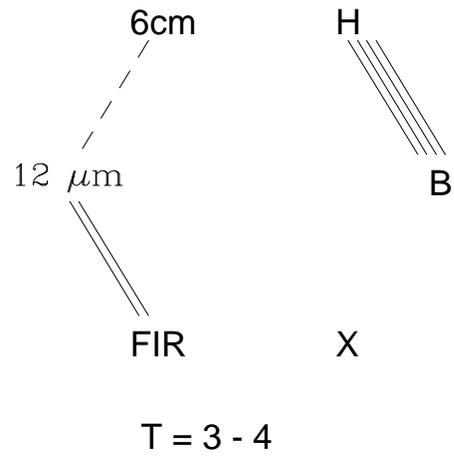
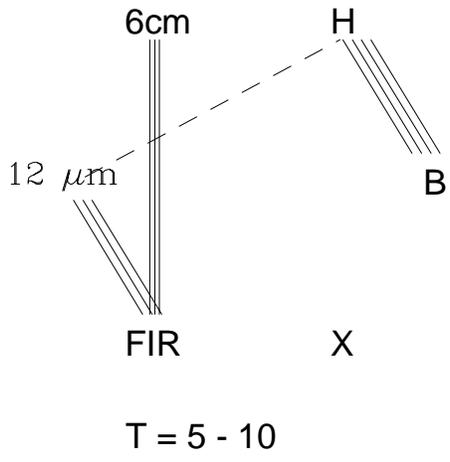
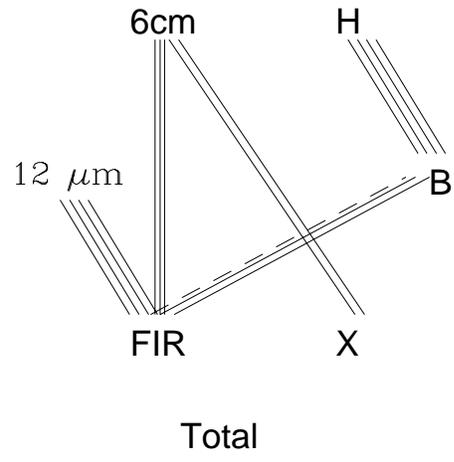
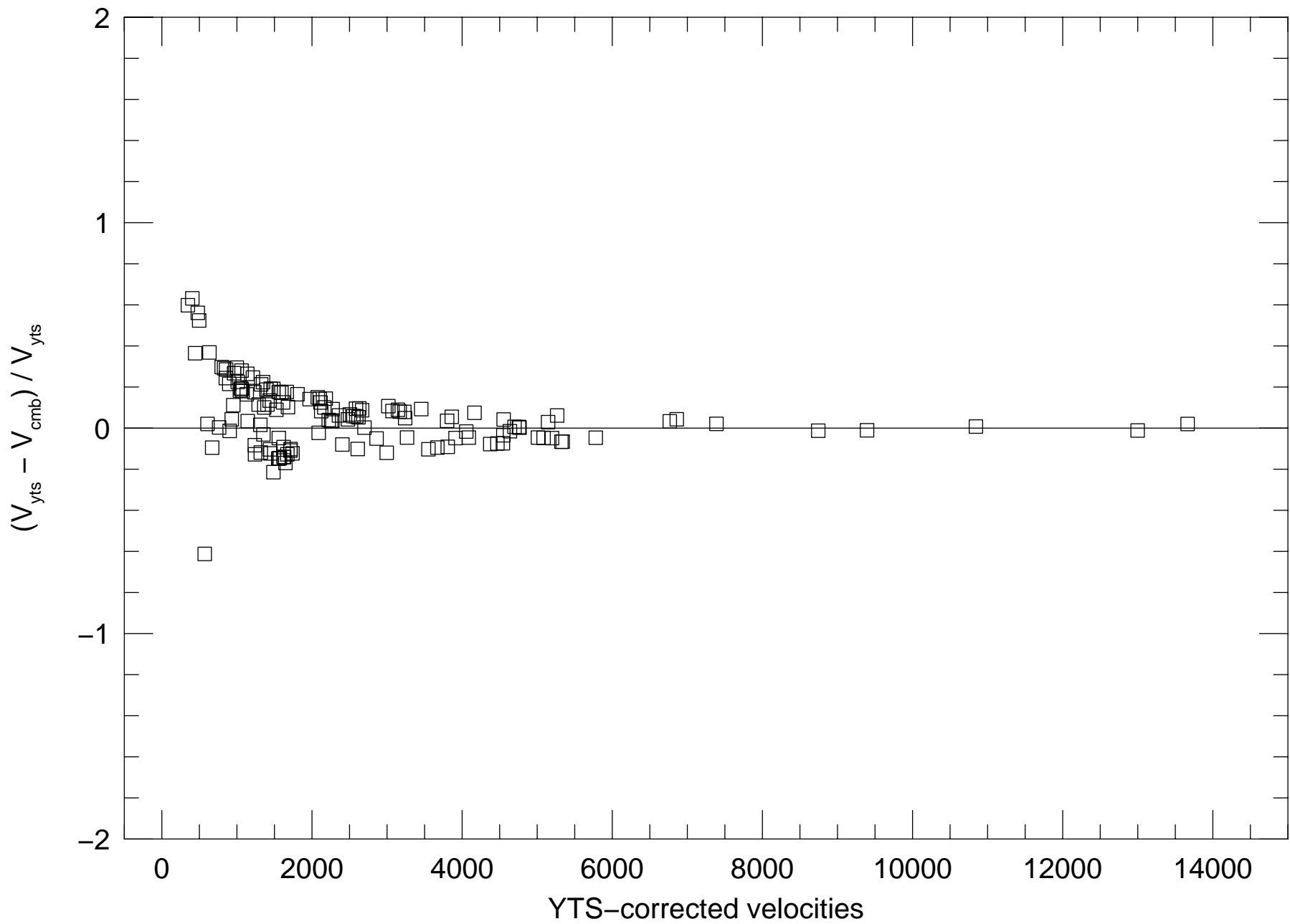
\end{document}